\documentclass{article}

\usepackage{graphicx}
\usepackage{amsmath,amssymb,amsfonts,theorem}
\usepackage{graphicx,epsfig,color,psfrag}
\usepackage{subfig}
\graphicspath{{./}}
\newcommand{\subscr}[2]{#1_{\textup{#2}}}
\newcommand{\supscr}[2]{#1^{\textup{#2}}}

\newcommand{\map}[3]{#1: #2 \rightarrow #3}

\newcommand{\degs}{^\circ\!}
\newcommand{\real}{\ensuremath{\mathbb{R}}}

\newcommand{\realnonnegative}{\ensuremath{\mathbb{R}}_{\geq0}}
\renewcommand{\natural}{{\mathbb{N}}}

\newcommand{\vc}{\supscr{v}{a}}
\newcommand{\vr}{\supscr{v}{r}}

\newcommand{\rreg}[2]{\mathcal{R}_{#1}^{#2}}
\newcommand{\creg}[2]{\mathcal{A}_{#1}^{#2}}

\newcommand{\vmax}{\subscr{v}{max}}

\newcommand{\alphac}{\subscr{\alpha}{a}}
\newcommand{\alphar}{\subscr{\alpha}{r}}
\newcommand{\Rsr}{\subscr{R}{sr}}
\newcommand{\fc}{\subscr{f}{a}}
\newcommand{\fr}{\subscr{f}{r}}
\newcommand{\Fc}{\subscr{F}{a}}
\newcommand{\Fr}{\subscr{F}{r}}

\newtheorem{proposition}{Proposition}

\begin{document}

\title{On stability of V-like Formations}
\author{Max D. Steel}
\maketitle

\section{Introduction}
Group behavior has received much attention as a test case of self-organization.  There has been much written in recent years to investigate interactions within groups of agents.  These agents can be animals moving in an interactive way, such as birds, but can also refer to situations such as people driving in traffic.  The models that describe these interactions are able to reproduce different structures and patterns relating to the movement and interaction of the agents involved.  The advantages and necessities of this type of analysis in any complex biological, technological, economic, or social system are important and far-reaching.  Each model that we will discuss describes interaction between agents.\\
Focusing on animal groups, such systems received 
attention not only from biologists and ethologists, but also from physicists, mathematicians, and engineers. This interest has produced a huge amount of literature, which is well documented and reviewed: see, for instance, \cite{IG:08,JK-GDR:02,DS:06} and~\cite{FB-JC-SM:09}. Loosely speaking, the basic idea behind these works is that complex collective behavior arises from simple interactions among close animals.

This paper considers a simple case of five birds in V formation, flying at constant velocity.
The aim is to study the feasibility of this solution and its stability for three models proposed in the literature.\\
The first model we will analyze was proposed in \cite{CFP}. It is an \textit{agent-based model}, in the sense that each agent is singularly considered.  The model is also \textit{leaderless}, meaning that all agents behave following the same set of rules and their behavior is not imposed by others.  The model is purely based on attraction-repulsion intractions between group mates.  Attraction allows the group to be formed and stay tight, while repulsion allows for coliision avoidance between group mates and keeps them well spaced.  This model does not combine attraction and repulsion with velocity alignment, unlike many other models that have been proposed.  The model also assumes that each agent interacts with a limited number of group mates that are in a suitable pre-defined sensitivity zone.  The authors were able to produce cluster formations, line formations and V-like formations by altering the attraction and repulsion parameters of their model.\\
The second model was proposed by Cucker and Smale in \cite{CS} and additional results are available in \cite{S}.
The model is of second order, in the sense that the dynamics is given for position and velocity of each animal.
More precisely, each animal adjusts its velocity to the mating ones, which are in its "interaction network".
The adjustment is done towards an average value of the various different velocities. Therefore this approach
is quite related to the theory of consensus.\\
The third model is again a consensus-type model proposed by Olfati-Saber in \cite{OS}.
In this case, beside the adjustment of velocity towards an average of mating birds, there is a term
which forces animal to keep a desired distance $d$. The latter is thought as a suitable distance which
taken into account both attraction among birds and repulsion for avoidance of conflicting trajectories.

For each model, we first show that the V formation is an admissible solution (for the first model
we need to accurately choose parameters). Then we study stability of the linearization along
the given V formation trajectory. The result is that the linearization of the first model
gives rise to negative or zero eigenvalues, while the other models presents also positive eigenvalues.
We can conclude that the second and third model are exponentially unstable.\\
In other words, despite the fact that these models can represent V like formations, they are not
able to represent stable V like formations. This indicates the need of further modelling efforts.

\section{Animal groups models}

Let us start with the model proposed in \cite{CFP}, which we refer to as the CFP model.\\
The animals in the model are represented by point particles, which have simple continuous-time dynamics.
Given $N\in\natural$, for all $i=1,\dots,N$ and $t\in \realnonnegative$, let $x_i(t)\in \real^2$ 
represent the position of the $i$-th animal, whose evolution is described by the differential equation
\begin{equation}\label{ode_fondamentale}
\dot x_i(t)=v_i(x(t)),
\end{equation}
where $x(t)$ is the vector $(x_1(t),\ldots,x_N(t))$. \\

The velocity $v_i(x)$ is the sum of two contributions, expressing the effects of attraction and repulsion,
$$v_i(x)=\vc_i(x)-\vr_i(x).$$
In more detail, each of these contributions depends on the relative position of the other animals,
\begin{align*}
\vc_i(x)=&\sum_{j\in \creg{i}{n}}\fc(\|x_j-x_i\|)\frac{x_j-x_i}{\|x_j-x_i\|}\\
\vr_i(x)=&\sum_{j\in \rreg{i}{n}}\fr(\|x_j-x_i\|)\frac{x_j-x_i}{\|x_j-x_i\|}.
\end{align*}
The following definitions have been used.
\begin{itemize}
\item The function $\map{\fc}{\realnonnegative}{\realnonnegative}$  (resp., $\map{\fr}{\realnonnegative}{\realnonnegative}$) describes how each animal is attracted (resp., repelled) by a neighbor at a given distance, assuming $\|\cdot\|$ denotes the Euclidean norm in $\real^2$.
\item The {\em attraction neighborhood} $\creg{i}{n}$ (resp., the {\em repulsion neighborhood} $\rreg{i}{n}$) is the set of the $n$ animals closest to the $i$-th one, which are inside the attraction (resp., repulsion) sensitivity zone.
\end{itemize}

\begin{figure}[!ht]\centering
\includegraphics[width=.6\columnwidth]{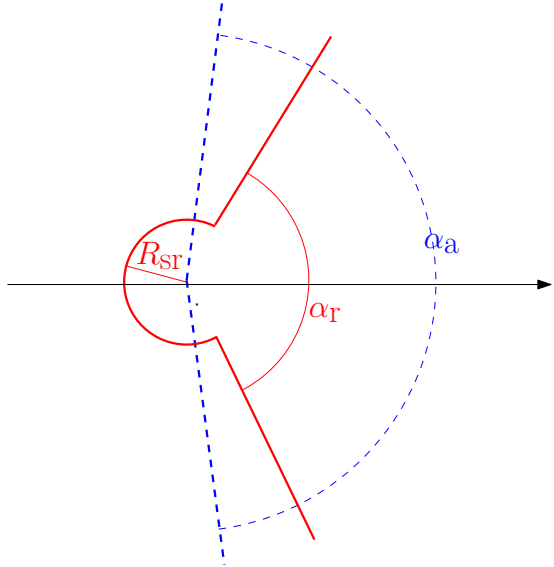}
  \caption{The shape of the sensitivity zones.} \label{fig:regions}
\end{figure}

The above model is very general, and we need to specialize it by choosing the interaction functions and the shape of the sensitivity zones. We make the following assumptions.
\begin{enumerate}
\item[A1.] The functions $\fc$ and $\fr$ are assumed to be 
$$ 
\fc(\|x_j-x_i\|)= \Fc \|x_j-x_i\|\,, \qquad \fr(\|x_j-x_i\|)=\frac{\Fr}{\|x_j-x_i\|},
$$ 
where $\Fc$ and $\Fr$ are two positive constants.
\item[A2.] The sensitivity zones are depicted in Figure~\ref{fig:regions} and illustrated as follows. Let the center point be the animal's position, and let the horizontal axis (arrow-headed) represent the direction of motion. Attraction is active in a frontal cone whose width is given by the angle $\alphac\in(0,360\degs]$ (dashed line). Repulsion is active both inside a disk of radius $\Rsr>0$ (\emph{short-range} repulsion) and in a frontal cone of width $\alphar\in(0\degs,360\degs]$ (solid line). We stress that the sensitivity zones do not necessarily coincide with the visual field of the animal. They rather represent the zones which attraction and repulsion are focused on.
\item[A3.] The speed of each animal $\|v_i\|$ is bounded from above by a constant $\vmax.$
\end{enumerate}

The above assumptions result in the system
\begin{equation}\label{eq:system-simulated}
\dot x_i(t)=\Fc\sum_{j\in \creg{i}{n}} (x_j-x_i)  -  \Fr \sum_{j\in \rreg{i}{n}} \frac{(x_j-x_i)}{\|x_j-x_i\|^2} .
\end{equation}
Some remarks are in order.
\begin{enumerate}
\item[R1.] The definition of the interaction neighborhoods $\creg{i}{n}$ and $\rreg{i}{n}$ allows to have an \textit{a priori} bound on the number of effective neighbors, and therefore on the sensing and ``computational'' effort which is required for each animal. This fact, which copes with animals' intrinsic limitations, has been experimentally observed in biology, for fish~\cite{IA:80} and birds~\cite{AC-IG:08a}. The latter paper calls this neighborhood definition {\em topological}, as opposed to {\em metric} definitions, based on distance only.
\item[R2.] Our assumption of unbounded sensitivity regions does not intend to imply that animals sensing capabilities extend on an unlimited range, but rather that group dynamics happen in a relatively small area.
\item[R3.] If $\alphac=\alphar=360\degs$, the parameter $\Rsr$ has no effect, and the interaction is completely isotropic as in the simulations presented in~\cite{AC-IG:08a,DC-SAL:99}. Note that, even if there is no preference for any specific direction, the limitation of the number of considered neighbors makes the interactions not reciprocal, i.e. the fact that the $i$-th animal interacts with the $j$-th does not imply that the $j$-th interacts with the $i$-th.
\item[R4.] By specializing the functions $\fc$ and $\fr$, one can obtain various interaction models. 
For instance, it is natural to assume that $\fr$ be decreasing and $\lim_{s\to \infty}{\fr(s)}=0$.
Similar considerations are valid also for some features introduced in other models, such as a \textit{neutral zone} around animals 
or a hierarchical decision tree which allows the repulsion force to have the priority over the attraction force.
\end{enumerate}

\vskip 5mm

The second model we will analyze is the Cucker-Smale model.
It is inspired by the Cucker-Smale flocking model, and attempts to analyze animal behavior under hierarchical leadership.  
The goal of this flocking study is to be able to interpret, model, analyze, predict, and simulate various flocking or multi-agent behavior.  The Cucker-Smale flocking model includes an interaction weight function which will allow us to analyze the birds in our formation by defining the weight function appropriately.  In a model of hierarchical leadership, the agents are ranked in some way and are therefore not independently motivated in their movement - that is, lower-rank agents are led by agents of higher ranks.  Using these assumptions, the paper goes on to discuss various flocks using these assumptions.\\
The Cucker-Smale flocking model which is given by the autonomous dynamical system:
\[
\left\{\begin{array}{ll}\dot x_i(t)=v_i, \\
\dot v_i(t)=\sum_{j\in L(i)} a_{ij}(x)(v_j-v_i),\qquad i = 1,\ldots,k,\ t > 0\end{array}\right.
\]
This model differs from the previous model in that the motion in each bird is defined in terms of $\dot x_i(t)$ and $\dot v_i(t)$.
In other words, this is a second order model, assuming acceleration driven by "social forces".
In this case $L(i)$ denotes the subgroup of animals that directly influence agent $i$.  

\vskip 5mm

The third model we will analyze is from \cite{OS}. It presents a theoretical framework for design and analysis of distributed flocking algorithms.  In this paper, three flocking algorithms are presented:  two are for free-flocking and one for constrained flocking.  
As in the Cucker-Smale model, this model compensates for velocity alignment.  
Their first algorithm generally leads to fragmentation, whereas the second and third algorithms both lead to flocking behavior.
However, unlike the second model, this paper asserts that flocks do not need leaders.  
The paper also includes various simulation results that demonstrate the effectiveness of the algorithms and analytical tools contained therein.\\  
The model has a format similar to the Cucker-Smale ones.  The motion of each bird's velocity described as follows:\\
\[
\dot v_i(t)=\sum_{j\in L(i)} \frac{v_j-v_i)}{\|x_j-x_i\|^2} + \sum_{j\in L(i)}(\|x_j-x_i\|^2 - d)(x_j-x_i)
\]

\section{V formation}
For purposes of this paper, we will consider five birds flying in a V-like formation in the context of three different models that have been put forth by various authors.  We will consider the motion of the birds over time and determine if equilibrium will be attained, and if that will lead to long-term stability, based on these models.\\

We will assume that the birds fly in a coordinate plane parallel to the ground.  Further, we will consider five birds $x_1$,..., $x_5$, and we will define their initial positions as follows:\\
\[
x_1 = (-2, 2),\quad  
x_2 = (-1, 1),\quad
x_3 = (0, 0), \quad
x_4 = (-1, -1)\quad
x_5 = (-2, -2).
\]
Further, we will assume that the initial velocity of the birds is translation to the right at velocity (1, 0) so they satisfy:\\
\[
x_i(t) = x_{i, 0} + (t, 0)\qquad	where\quad i = 1, ..., 5
\]
We will call this the V evolution.\\

\subsection{CFP model}
As stated above, this model considers a force of attraction and a force of repulsion.  Each animal is both attracted and repulsed by the other birds by predetermined parameters.\\ 
\\
We want the V evolution to solve the equations as described below:
\[
\dot x_i(t)=F_a\sum_{j\in \creg{i}{n}} (x_j-x_i)\:-\:F_r\sum_{j\in \rreg{i}{n}} \frac{(x_j-x_i)}{\|x_j-x_i\|^2} .
\]\\
In these equations, $F_a$ and $F_r$ are two positive constants.  As we will be setting $\dot x_i(t)$ to zero, we will need to determine $F_a$ and $F_r$ explicity.\\
The region in the last equation depends on parameters $\alpha _R$(angle of anisotropic repulsion regions), $R_{sr}$ (radius of circular repulsion region) and $\alpha _A$ (angle of anisotropic attraction region).\\

We will consider two cases:
\begin{itemize}
\item[case 1] $\alpha_r=90$, $R_{sr}=\sqrt{2}$ and $\alpha_a=179$.
\item[case 2] $\alpha_r=360$, $R_{sr}=\sqrt{2}$ and $\alpha_a=360$.
\end{itemize}
Let us first address case 1.  
Then, animal $x_3$ will not consider the other birds, and will therefore keep constant velocity of (1, 0).  Let us consider $x_4$.  This animal will feel both attraction and repulsion towards $x_3$, and not for the other birds.\\
Therefore, the equation for the motion of $x_4$ will be defined as follows:
\[
\dot x_4(t)=\Fc (x_3-x_4)  -  \Fr \frac{(x_3-x_4)}{\|x_3-x_4\|^2} .
\]
For $x_4$ above to be a solution to the model, we need the right-hand side of the equation to be zero.  Taking into account that ($x_3$, $x_4$) = (1, -1) this means
\[
F_a(1,1)  - \frac{F_r}{2} (1,1)=0,
\]
thus we need 2$F_a$=$F_r$.\\
The evolution for each bird in the V formation can be described as:
\[
\dot x_i = \varphi _i(x_1, x_2, ... ,x_5). 
\]
where we define
\[
X = \left[\begin{array}{c}x_1 \\ x_2\\  x_3 \\ x_4 \\ x_5\end{array}\right] = 
\left[\begin{array}{c}x_{1, 1} \\ x_{1, 2} \\ x_{2, 1} \\ x_{2,2}  \\ ... \\ x_{5,1} \\ x_{5,2}\end{array}\right]\\
\]\\
The evolution of the whole formation can be written as
\[ \dot X= F(X),\]
where
\[
F= \left[\begin{array}{c}\varphi_1 \\ \varphi_2 \\ \varphi_3 \\ \varphi_4 \\ \varphi_5\end{array}\right] 
= \left[\begin{array}{c}F_1 \\ F_2 \\ F_3 \\ ... \\ F_{10}\end{array}\right]\\
\]\\

We will linearize our system around our solution, the V solution.  We will compute the Jacobian matrix of the vector F of the right hand side of the equation above, for $i = 1, \dots, 5$ along the V solution. 
Thus F has a total of ten components (two for each bird) with each element of the matrix determined 
by the partial derivatives of F with respect to each component of X as:
\[
\frac{\partial F_j}{\partial x_{i,j}}
\]
For instance, the two rows of the matrix that define the motion of $x_4$ are $F_7$ and $F_8$.\\
It follows that
\[
\frac{\partial F_7}{\partial x_{3,1}}\:=\:
F_a-F_r\frac{(x_{32}-x_{42})^2-(x_{31}-x_{41})^2)}{\parallel x_3-x_4 \parallel^4}
\]
This leads to the value of $\frac{1}{2}$, which is inserted in the matrix.   We then compute the other values for $F_7$ and $F_8$ using partial derivatives.  We perform similar calculations for birds $x_1$, $x_2$, and $x_5$, assuming that $x_1$ considers birds $x_2$ and $x_3$, $x_2$ only considers $x_3$ and $x_5$ considers $x_3$ and $x_4$.  This leads to the following matrix:
\[
\left[\begin{array}{cccccccccc}
	-1 & -0.5 & 0.5 & 0.5 & 0.5 & 0 & 0 & 0 & 0 & 0\\
	-0.5 & -1 & 0.5 & 0.5 & 0 & 0.5 & 0 & 0 & 0 & 0\\
	0 & 0 & -0.5 & 0.5 & 0.5 & -0.5 & 0 & 0 & 0 & 0\\
	0 & 0 & 0.5 & -0.5 & -0.5 & 0.5 & 0 & 0 & 0 & 0\\
	0 & 0 & 0 & 0 & 0 & 0 & 0 & 0 & 0 & 0\\
	0 & 0 & 0 & 0 & 0 & 0 & 0 & 0 & 0 & 0\\
	0 & 0 & 0 & 0 & 0.5 & 0.5 & -0.5 & -0.5 & 0 & 0\\
	0 & 0 & 0 & 0 & 0.5 & 0.5 & 0.5 & -0.5 & 0 & 0\\
	0 & 0 & 0 & 0 & 0.5 & 0 & 0.5 & -0.5 & -1 & 0.5\\
	0 & 0 & 0 & 0 & 0 & 0.5 & -0.5 & 0.5 & 0.5 & -1
	\end{array}
\right]
\]

This matrix has eigenvalues of 0, -0.5, -1, -1.5.  We can thus write:
\begin{proposition}
Consider the CFP model with $\alpha_r=90$, $R_{sr}=\sqrt{2}$ and $\alpha_a=179$.
Then the V formation is admissible and not exponentially unstable.
\end{proposition}

Let us now address case 2.  In this case, $x_3$ feels both attraction and repulsion 
for $x_2$ and $x_3$ (since we are now considering a circular repulsion region).  This leads to the following equation:
\[
\dot x_3(t)=F_a ((x_2-x_3)+(x_4-x_3))  -  F_r (\frac{(x_2-x_3)}{\|x_2-x_3\|^2} + \frac{(x_4-x_3)}{\|x_4-x_3\|^2}).
\]
However, the model fails to admit V formation as solution.
When attempting to determine the coefficients for $F_a$ and $F_r$, it becomes impossible to determine 
constants that will allow the right side of the equation above to be set equal to zero.

\subsection{The Cucker-Smale model}
For this model, we will apply similar computations and construct a similar matrix to the last model, 
this matrix will now be 20 by 20 instead of 10 by 10, as each bird will have four components in the matrix, 
two for $x_i$(t) and two for $v_i$(t).\\
Since $x_i$(t) = $v_i$ for each bird, the rows of each matrix that correspond with $\dot x_i$ will take the following form in the matrix:
\[
\left[\begin{array}{cccc}
0 & 0 & 1 & 0\\
0 & 0 & 0 & 1
\end{array}\right]
\]
when computing the partial derivatives with the four columns $x_{i1}$, $x_{i2}$, $v_{i1}$, and $v_{i2}$.
The rows that correspond to $\dot v_i$(t) will be computed with partial derivatives as before.  The form for each $\dot v_i$(t) will look as follows:
\[
V_i = \frac {1}{\|x_i-x_j|^2}(v_j-v_i)
\]
for the bird(s) appropriately chosen.

We will assume that each bird interacts with the other birds as in Case 1 for the CFP model.  
That is, bird 3 is independent of the other birds and therefore produces zeroes in the matrix for the four 
rows corresponding to bird 3.  We compute partial derivatives as in the first model, and this generates the following matrix:

{\scriptsize
\[
\left[\begin{array}{cccccccccccccccccccc}
	0 & 0 & 1 & 0 & 0 & 0 & 0 & 0 & 0 & 0 & 0 & 0 & 0 & 0 & 0 & 0 & 0 & 0 & 0 & 0\\
 	0 & 0 & 0 & 1 & 0 & 0 & 0 & 0 & 0 & 0 & 0 & 0 & 0 & 0 & 0 & 0 & 0 & 0 & 0 & 0\\
	0 & 0 & -1 & 0 & 0 & 0 & 1 & 0 & 0 & 0 & 0 & 0 & 0 & 0 & 0 & 0 & 0 & 0 & 0 & 0\\
	0 & 0 & 0 & -1 & 0 & 0 & 0 & 1 & 0 & 0 & 0 & 0 & 0 & 0 & 0 & 0 & 0 & 0 & 0 & 0\\
	0 & 0 & 0 & 0 & 0 & 0 & 1 & 0 & 0 & 0 & 0 & 0 & 0 & 0 & 0 & 0 & 0 & 0 & 0 & 0\\
	0 & 0 & 0 & 0 & 0 & 0 & 0 & 1 & 0 & 0 & 0 & 0 & 0 & 0 & 0 & 0 & 0 & 0 & 0 & 0\\
	0 & 0 & 0 & 0 & 0 & 0 & -1 & 0 & 0 & 0 & 1 & 0 & 0 & 0 & 0 & 0 & 0 & 0 & 0 & 0\\
	0 & 0 & 0 & 0 & 0 & 0 & 0 & -1 & 0 & 0 & 0 & 1 & 0 & 0 & 0 & 0 & 0 & 0 & 0 & 0\\
 	0 & 0 & 0 & 0 & 0 & 0 & 0 & 0 & 0 & 0 & 1 & 0 & 0 & 0 & 0 & 0 & 0 & 0 & 0 & 0\\
	0 & 0 & 0 & 0 & 0 & 0 & 0 & 0 & 0 & 0 & 0 & 1 & 0 & 0 & 0 & 0 & 0 & 0 & 0 & 0\\
	0 & 0 & 0 & 0 & 0 & 0 & 0 & 0 & 0 & 0 & 0 & 0 & 0 & 0 & 0 & 0 & 0 & 0 & 0 & 0\\
	0 & 0 & 0 & 0 & 0 & 0 & 0 & 0 & 0 & 0 & 0 & 0 & 0 & 0 & 0 & 0 & 0 & 0 & 0 & 0\\
	0 & 0 & 0 & 0 & 0 & 0 & 0 & 0 & 0 & 0 & 0 & 0 & 0 & 0 & 1 & 0 & 0 & 0 & 0 & 0\\
	0 & 0 & 0 & 0 & 0 & 0 & 0 & 0 & 0 & 0 & 0 & 0 & 0 & 0 & 0 & 1 & 0 & 0 & 0 & 0\\
	0 & 0 & 0 & 0 & 0 & 0 & 0 & 0 & 0 & 0 & -1 & 0 & 0 & 0 & 1 & 0 & 0 & 0 & 0 & 0\\
	0 & 0 & 0 & 0 & 0 & 0 & 0 & 0 & 0 & 0 & 0 & -1 & 0 & 0 & 0 & 1 & 0 & 0 & 0 & 0\\
	0 & 0 & 0 & 0 & 0 & 0 & 0 & 0 & 0 & 0 & 0 & 0 & 0 & 0 & 0 & 0 & 0 & 0 & 1 & 0\\
	0 & 0 & 0 & 0 & 0 & 0 & 0 & 0 & 0 & 0 & 0 & 0 & 0 & 0 & 0 & 0 & 0 & 0 & 0 & 1\\
	0 & 0 & 0 & 0 & 0 & 0 & 0 & 0 & 0 & 0 & 0 & 0 & 0 & 0 & -1 & 0 & 0 & 0 & 1 & 0\\
	0 & 0 & 0 & 0 & 0 & 0 & 0 & 0 & 0 & 0 & 0 & 0 & 0 & 0 & 0 & -1 & 0 & 0 & 0 & 1
	\end{array}
\right]
\]
}\\

This matrix has eigenvalues of 0, 1, and -1.  This suggests that the birds will not attain stability or equilibrium in a simulation of the model.\\

\begin{proposition}
Consider the Cucker-Smale model.
Then the V formation is admissible and is exponentially unstable.
\end{proposition}

\section{The Olfati-Saber model}
This model is quite similar to the Cucker-Smale ones, thus we will only indicate the main differences.
The sum of terms in the derivative of velocity, is only taken for birds interacting, so we will consider the birds interacting as in 
Cucker-Smale model ($x_1$ only considers $x_2$, $x_2$ only considers $x_3$, $x_3$ is independent, etc.)

We will define d = $\sqrt{2}$ so that the birds will interact as described before.  We compute partial derivatives as before and 
compute the following matrix:

$\hskip -4.5cm{\scriptsize
 \left[\begin{array}{cccccccccccccccccccc}
	0 & 0 & 1 & 0 & 0 & 0 & 0 & 0 & 0 & 0 & 0 & 0 & 0 & 0 & 0 & 0 & 0 & 0 & 0 & 0\\
 	0 & 0 & 0 & 1 & 0 & 0 & 0 & 0 & 0 & 0 & 0 & 0 & 0 & 0 & 0 & 0 & 0 & 0 & 0 & 0\\
	\sqrt{2} - 4 & -2 & 0 & 4 - \sqrt{2} & -2 & 2 & 0 & 0 & 0 & 0 & 0 & 0 & 0 & 0 & 0 & 0 & 0 & 0 & 0 & 0\\
	2 & \sqrt{2}-4 & 0 & -2 & -2 & 4 - \sqrt{2} & 0 & 2 & 0 & 0 & 0 & 0 & 0 & 0 & 0 & 0 & 0 & 0 & 0 & 0\\
	0 & 0 & 0 & 0 & 0 & 0 & 1 & 0 & 0 & 0 & 0 & 0 & 0 & 0 & 0 & 0 & 0 & 0 & 0 & 0\\
	0 & 0 & 0 & 0 & 0 & 0 & 0 & 1 & 0 & 0 & 0 & 0 & 0 & 0 & 0 & 0 & 0 & 0 & 0 & 0\\
	0 & 0 & 0 & 0 & \sqrt{2}-4 & 2 & -2 & 0 & 4 -\sqrt{2} & -2 & 2 & 0 & 0 & 0 & 0 & 0 & 0 & 0 & 0 & 0\\
	0 & 0 & 0 & 0 & 2 & \sqrt{2}-4 & 0 & -2 & -2 & 4-\sqrt{2} & 0 & 2 & 0 & 0 & 0 & 0 & 0 & 0 & 0 & 0\\
 	0 & 0 & 0 & 0 & 0 & 0 & 0 & 0 & 0 & 0 & 0 & 0 & 0 & 0 & 0 & 0 & 0 & 0 & 0 & 0\\
	0 & 0 & 0 & 0 & 0 & 0 & 0 & 0 & 0 & 0 & 0 & 0 & 0 & 0 & 0 & 0 & 0 & 0 & 0 & 0\\
	0 & 0 & 0 & 0 & 0 & 0 & 0 & 0 & 0 & 0 & 0 & 0 & 0 & 0 & 0 & 0 & 0 & 0 & 0 & 0\\
	0 & 0 & 0 & 0 & 0 & 0 & 0 & 0 & 0 & 0 & 0 & 0 & 0 & 0 & 0 & 0 & 0 & 0 & 0 & 0\\
	0 & 0 & 0 & 0 & 0 & 0 & 0 & 0 & 0 & 0 & 0 & 0 & 0 & 0 & 1 & 0 & 0 & 0 & 0 & 0\\
	0 & 0 & 0 & 0 & 0 & 0 & 0 & 0 & 0 & 0 & 0 & 0 & 0 & 0 & 0 & 1 & 0 & 0 & 0 & 0\\
	0 & 0 & 0 & 0 & 0 & 0 & 0 & 0 & \sqrt{2}-4 & 2 & -2 & 0 & 4-\sqrt{2} & -2 & 2 & 0 & 0 & 0 & 0 & 0\\
	0 & 0 & 0 & 0 & 0 & 0 & 0 & 0 & 2 & \sqrt{2}-4 & 0 & -2 & -2 & 4-\sqrt{2} & 0 & 2 & 0 & 0 & 0 & 0\\
	0 & 0 & 0 & 0 & 0 & 0 & 0 & 0 & 0 & 0 & 0 & 0 & 0 & 0 & 0 & 0 & 0 & 0 & 1 & 0\\
	0 & 0 & 0 & 0 & 0 & 0 & 0 & 0 & 0 & 0 & 0 & 0 & 0 & 0 & 0 & 0 & 0 & 0 & 0 & 1\\
	0 & 0 & 0 & 0 & 0 & 0 & 0 & 0 & 0 & 0 & 0 & 0 & \sqrt{2}-4 & 2 & -2 & 0 & 4-\sqrt{2} & -2 & 2 & 0\\
	0 & 0 & 0 & 0 & 0 & 0 & 0 & 0 & 0 & 0 & 0 & 0 & 2 & \sqrt{2}-4 & 0 & -2 & -2 & 4-\sqrt{2} & 0 & 2
	\end{array}
\right]}$

\vskip 5mm

This matrix generates eigenvalues of -1, -0.356, -1.644, -1.363, 3.363, 2.259 and 0. 
\begin{proposition}
Consider the Olfati-Saber model.
Then the V formation is admissible and is exponentially unstable.
\end{proposition}

\section{Conclusions}
We considered three recent models proposed to describe the behaviour of multi-agent systems, in particular animal groups.
A simple V like formation was tested on each model in terms of admissibility of evolution and stability.
The main result is that V formation are exponentially unstable for two models. This indicated the need
of further modeling efforts in order to appropriately address the evolution of birds flocks.

\end{document}